# A NOVEL APPROACH AGAINST E-MAIL ATTACKS DERIVED FROM USER-AWARENESS BASED TECHNIQUES

Gaurav Ojha[1] and Gaurav Kumar Tak[2]

[1]Department of Information Technology, Indian Institute of Information Technology and Management, Gwalior - 474010, India
`ojhagaurav@ymail.com`
[2]School of Computer Science & Information Technology, Lovely Professional University, Phagwara, Punjab – 144402, India
`gauravtakswm@gmail.com`

## ABSTRACT

*A large part of modern day communications are carried out through the medium of E-mails, especially corporate communications. More and more people are using E-mail for personal uses too. Companies also send notifications to their customers in E-mail. In fact, in the Multinational business scenario E-mail is the most convenient and sought-after method of communication. Important features of E-mail such as its speed, reliability, efficient storage options and a large number of added facilities make it highly popular among people from all sectors of business and society. But being largely popular has its negative aspects too. E-mails are the preferred medium for a large number of attacks over the internet. Some of the most popular attacks over the internet include spams, and phishing mails. Both spammers and phishers utilize E-mail services quite efficiently in spite of a large number of detection and prevention techniques already in place. Very few methods are actually good in detection/prevention of spam/phishing related mails but they have higher false positives. A variety of filters such as Bayesian filters, Checksum-based filters, machine learning based filters and memory-based filters are usually used in order to identify spams. These techniques are implemented at the server and in addition to giving higher number of false positives, they add to the processing load on the server. This paper outlines a novel approach to detect not only spam, but also scams, phishing and advertisement related mails. In this method, we overcome the limitations of server-side detection techniques by utilizing some intelligence on the part of users. Keywords parsing, token separation and knowledge bases are used in the background to detect almost all E-mail attacks. The proposed methodology, if implemented, can help protect E-mail users from almost all kinds of unwanted mails with enhanced efficiency, reduced number of false positives while not increasing the load on E-mail servers.*

## KEYWORDS

*Spams, Scams, Phishing, Social networking, Advertisements & Subscriptions*

## 1. INTRODUCTION

E-mail is probably the most convenient method of transferring messages electronically from one person to another, emerging from and going to any part of the world. E-mail connects each and every location on earth, eliminating geographical boundaries and bringing people closer. On the technical side, it involves a number of protocols, such as SMTP, POP, TCP/IP and so on, for taking messages from one mailbox to another. E-mail has been aptly defined as a method for exchanging digital information from one author (sender) to one or more recipients and it has become the standard medium of communication in various areas of life. It provides many attractive features by its virtue such as fast, simple and free access, global acceptance, support for instant messaging protocols, support for file attachments, etc. [3]

  1

E-mail becomes essential due to being the key for a large number of internet services such as social networking, newsletter subscriptions, etc. Here, in the first section, we introduce about the concept of E-mails. Then we go about techniques to detect E-mail attacks in the second section. In the third section, we put forth the approach in our proposed methodology which the user can practice to protect him from E-mail attacks.

The various uses of E-mail, by people from various age groups, are outlined below

1. **A general method of communication:** As a medium of communication, one of the most important purposes served by E-mail is that of eliminating distances and helping to keep in touch. Be it family or old friends or just about anyone, with E-mail nobody is ever away. Gone are the days when getting replies to letters meant weeks of waiting period. With E-mail, one can get a reply instantly and can even communicate in close to real-time. For students staying far away from their parents, nothing can serve as better as E-mail can. In this way, it becomes a general method of communication for everyone.

2. **Academic uses:** In many universities, E-mail serves as a medium to share the course material. Faculties send course material and relevant presentations, or documents through E-mail to all the students in major universities across the world. This makes it possible to receive study material and complete the course. The facility of attaching files opens an option also whereby study material such as presentations and videos can be sent seamlessly over E-mails within a matter of seconds.

3. **Business correspondence:** E-mail is usually considered as the official medium of communication in the corporate sector. Nowadays, liberal trade policy is becoming more and more acceptable and the export and import of goods to and from the foreign markets has become a frequent affair. This is where E-mail comes into the picture containing business and trade. Nowadays, most of the business-related communications are done via E-mails. All different kinds of deals, tenders, quotations, are communicated through mails.

4. **Subscriptions and services:** E-mail is used as a method of verification on various social networking and subscription related websites. Services offered by Facebook, Blogger, matrimonial sites such as bharatmatrimony.com, shaadi.com, educational services such as knowafest.com, pratiyogi.com, etc. require an E-mail address compulsorily. Later on, E-mail is used to properly manage subscriptions and notifications from these sites. Even with the advent of 'Open Authorization' wherein it is possible to login through one's social networking accounts such as Facebook or Twitter, the primary requirement continues to be an E-mail ID.

5. **Events, contests and fests:** A variety of online competitions, require university students to submit their projects via E-mail. Conferences, journals, and traditional publishers accept material through E-mail. This makes E-mail a very important method of communication for all extra-curricular and research-related activities.

Apart from the above mentioned uses, there are various other ways in which E-mail can be useful and is utilized. Because E-mail is such a fool-proof method of communication, and has an enormous outreach, it also becomes a good medium for sending a large number of non-beneficial E-mails to unsuspecting users. Spamming and sending mails with viruses, worms or other kind of malware, which waste both time and network resources namely memory and bandwidth, are the most popular misuses of E-mails. Some E-mails also aim at stealing personal information of users or gain unauthorized access to their accounts. Many attacks have adverse effects on the users in terms of monetary losses and mental disturbance.

The various negative uses of E-mail are outlined below

1. **Spam:** Spams are undesired mails, which are sent to one's Inbox, and are of no use of the recipient. They are sent on the network to increase the consumption of resources, i.e. in other words, increase network traffic. E-mail has been widely used to distribute spam mails to many people around the world. Spams are widespread and they are most common in old E-mail IDs. As usage of an E-mail ID increases, the number of spams being received also increases. Many studies have found that spams are a result of giving out E-mail IDs at the time of registration for various subscriptions and accounts. Some spams include just hyperlinks, while some include text. Some links may be shortened more than once making it difficult to identify the website they are pointing to.

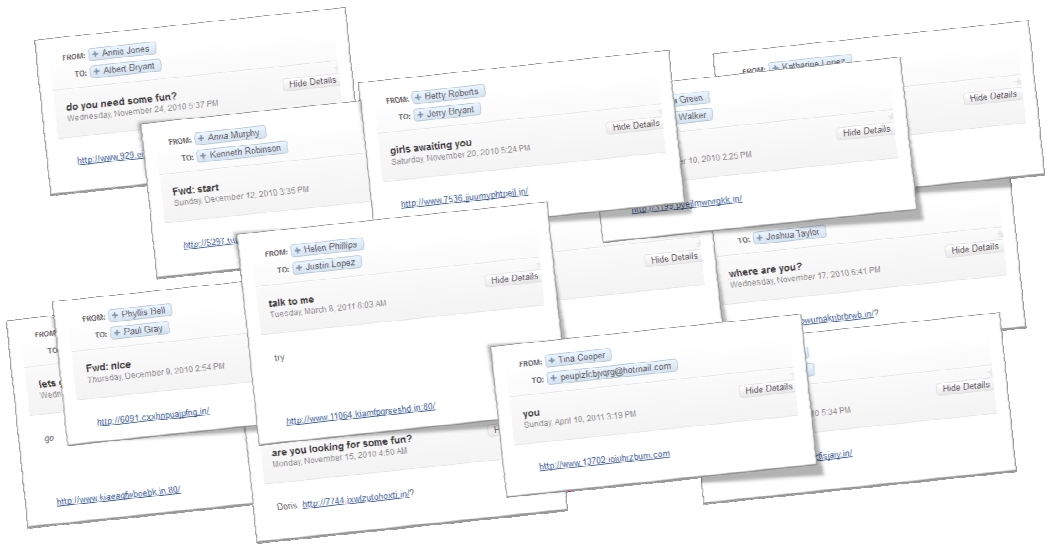

Figure 1. Examples of single-link spam E-mails (Destination address may be disguised)

2. **Scams:** Scams are a category of E-mails which are meant to lure a person into transactions which are bogus and completely fraudulent. It is believed that spams first began originating from Nigeria in the form of fake baking transactions. Any person falling prey to such fake transactions was sure to incur losses, and there was no one who could be blamed. The term 'Nigerian scam' was coined for such transactions, which began to get more and more sophisticated with time. Sometimes, they are combined with phishing. The attacker announces a lottery or any such prize of large monetary value, in return of users' personal information. Unaware users gladly provide such information only to be fooled into the trap. The attacker may then ask for some amount as transaction or processing charges for the prize amount, upon receiving which his purpose is served.

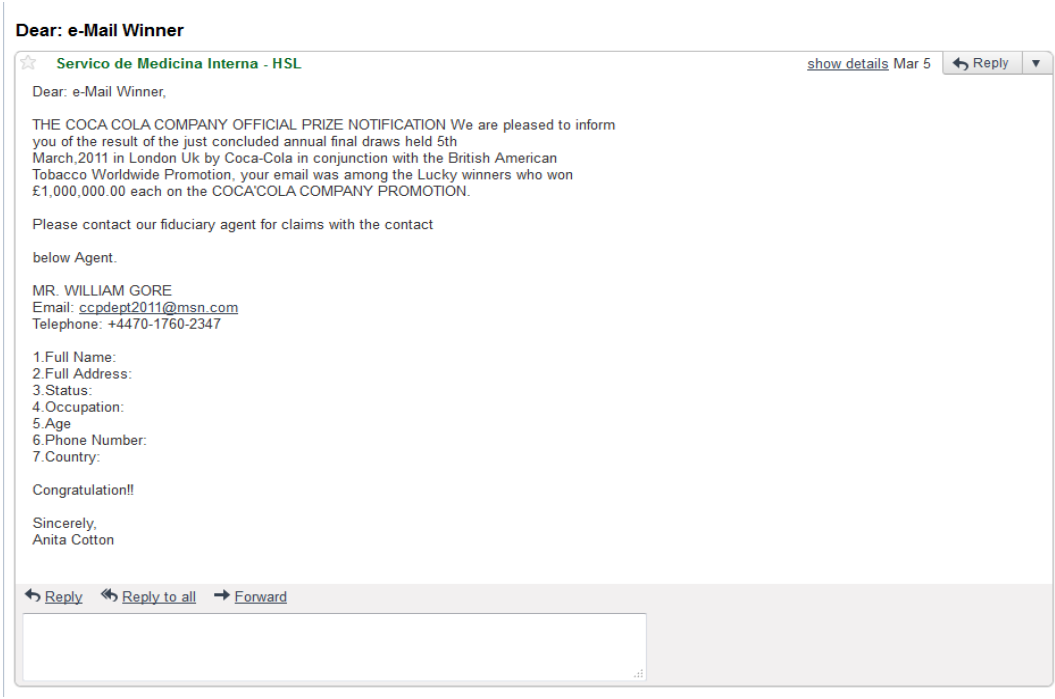

Figure 2. Example of a scam E-mail

3. **Spreading phishing URLs:** Phishing websites are fake websites, which are almost identical to some social networking websites, E-mail websites or Online Banking websites, etc., which require a user to login. In a phishing attack, a URL is registered and a website is uploaded, so that both the URL and the website look similar to an original website. The URL, now called a Phishing URL, is distributed across the internet using E-mail primarily. Phishing websites then capture the confidential login credentials of unsuspecting users, which may then be used to perform fraudulent transactions on their behalf. Sometimes the information may just be captured and used to perform some illegal activity in future.

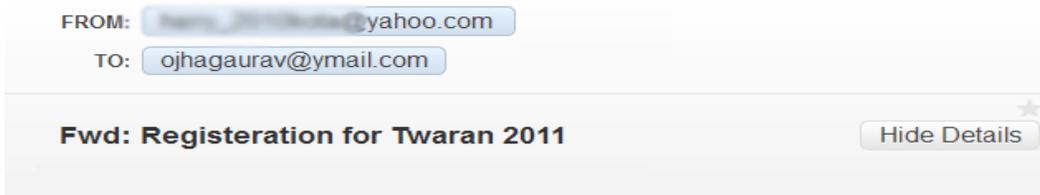

Figure 3. Example of a simple phishing URL sent in E-mail

4. **E-mail spoofing:** In spoofing, some important parts of the E-mail such as the header and the sender's address may be changed so as to imply a mail being sent by someone else. This kind of activity is possible as there is hardly any provision for authentication on SMTP, and changing crucial content becomes easy. It is usually a malicious activity, which is done to hide the origin of the mail, by changing some parts of the mail header. This is done in order to hide one's own identity. It is common for spamming and phishing. With spoofing, it is possible to misuse one's E-mail address in a variety of ways. Any kind of mail can be sent without the knowledge of the person who owns the E-mail ID. Ordinary PHP Scripts can do this job very perfectly. The simple 'mail()' function in PHP is all it takes to send mails. A simple script which can be used for spoofing is given below [18] [19]:

```
<?php
$to="recipient@example.com";
$subject="Login Details of your Amazon Account";
$from="admin@amazon.com";
$headers="From: $from";
if (mail($to, $subject, $body, $headers))
echo "<br/>E-mail spoofing done successfully";
else echo "<br/>E-mail spoofing failed";
?>
```

There are many other methods by which spoofing can be performed.

All these attacks are well-known and there are many more variations possible. Techniques have also been defined to prevent or detect these attacks and protect users from spammers and scammers, but all methods are not covered by these techniques. Some techniques require a lot of filtering functions and rely on complicated mathematical operations to separate spam from an ordinary mail. Although some techniques are efficient, they give a large number of false positives. Needless to mention, due to their resource-intensive operations, most of the techniques take up a lot of server resources in order to filter E-mail messages i.e., they have high memory, space and time complexities. So in this paper we shall discuss about a technique which shall emerge to be useful in such a scenario.

## 2. RELATED WORK

In literature, many discussions have described the techniques of detection of several types of spam or scams. Some have good efficiency but come up with a significant number of false positives. Some techniques are useful, but most of them are implemented on the server side. Some of the important works towards detection of E-mail attacks over the web are listed below.

In [2], the current scenario in spam and its detection has been highlighted. In [9], a rule-based approach has been implemented for the detection of spam mails, which is based on some learning process and intelligence. The discussed approach uses the training and testing phases of data for learning as well as for churning out better results. Moreover, the stale and obsolete spam rules are updated during the training phase. This action is used for improving the spam or scam filtering efficiency. However, the time as well as memory complexity is higher due to the generation of rules, managing data sets and a number of execution operations. E. Damiani et al. have described some basic properties and behaviour of spam mails. They have specified reasons of the popularity of spam mails and their results. The use of digests in the proposed approach to identify spam mails in a privacy-preserving way is a fundamental technique for collaborative learning [14]. A social network is designed based on the fact that they are used to exchange a lot of information as well as email address of registered users [11] [13]. Spammers are identified by observing misbehaviour or abnormalities in the structural properties of the network. Most of the times, email attackers use public social sites to harvest email addresses, (sometimes this task may even be performed by bots) to their mail list database. However, it is a reactive mechanism, which is used to identify spam mails and their behaviour since spammers are identified; in this approach, the spam filter uses previous history of spam mails. In [17], a novel and better approach has been described, which creates a Bayesian network out of email exchanges to detect spam. Though Bayesian classifiers are used for detecting spam emails, they inherently need to scan the contents of the email completely to compute the probability distributions for every node, which is present in the network, because many times, it is not possible to detect spam mails for a particular inbox and its requirement for filtering spam mails [5].

Nitin Jindal et al. discussed an approach of review on content-based spam mails. Review spam is quite different from Web page spam and E-mail spam, and thus requires different detection techniques [10]. Shashikant et al. also proposed separation techniques based on token separation of E-mail contents and some probabilistic approach. The proposed techniques provide very less false positive results. Firstly, the email server receives the E-mail content, and then it separates the tokens of E-mail content using some proper operations and analyses the content based on the requirement or need of users [1]. In [8] [12] [15] [16], some of these filters have been discussed to improve the spam detection and to protect the E-mails users against email attacks. Examples of filtering processes are Checksum based filtering, Bayesian spam filtering, Machine learning based classification, and Memory based filtering. In [3] [4], some techniques have been discussed, which focus on the partial match of attack keywords or spam keywords or spam content. In [3], the proposed approach uses some learning processes to identify E-mail attacks. One effective technique has already been proposed to identify the spam mail viz. 'Fast Effective Botnet Spam Detection'. It uses the header information of mails to detect the spam mails or other email attacks. It effectively works for both types of spam mails namely, 'Text based spam' as well as 'Image based spam'. It locates the sender's IP address, sender's email address; MX records and MX hosts, and analyses all the recorded information to provide effective results [7].

Some of the previous researches or discussions were based on the detection of spam or identification of scam and their filtering. In [6], Yan Luo has provided insights into various kinds of spam-filtering techniques, their space and time complexities and how they are implemented. But currently, several email attacks are not related to scam or spam attacks. In that situation we are proposing some new techniques to identify such email attacks, which provides efficient results with significantly small number of false positives.

## 3. PROPOSED METHODOLOGY

In various studies it has been found that most of the E-mail attacks happen because the users have subscriptions to various websites including social networking websites. Strictly speaking, most spam mails are a consequence of distributing E-mail IDs, knowingly or unknowingly, for various subscriptions and registration of online accounts. Since we have identified the source of a majority of spam and scam mails, we move ahead to suggest the following methodology, which can protect E-mail users from almost all kinds of E-mail related attacks.

### 3.1. Sharing of E-mail IDs with untrustworthy people or applications must be avoided

A large number of E-mail users casually distribute their E-mail IDs among friends, relatives and other people concerned. They may or may not be communicating regularly with these people, but they give their IDs irrespectively. It is possible that some people may try to hack their mailbox in order to capture personal or confidential information. Sometimes, their E-mail IDs maybe, intentionally or unintentionally, given to websites for recommending certain services. This might then result in various spam or pornographic advertisements arriving at one's mailbox. Some organizations even sell their contact list to various advertisement companies creating yet another source of spams. Spoofing is also possible when an untrustworthy party knows one's E-mail address. Many a times, users miss their important mails or messages because of spams or mails that are partially or completely avoidable.

Social networking websites usually provide a feature to invite friends by synchronizing contacts from one's E-mail account. Usage of this feature results in a number of unwanted/spam mails arriving at one's mailbox. See Fig. 4 and Fig. 5 below.

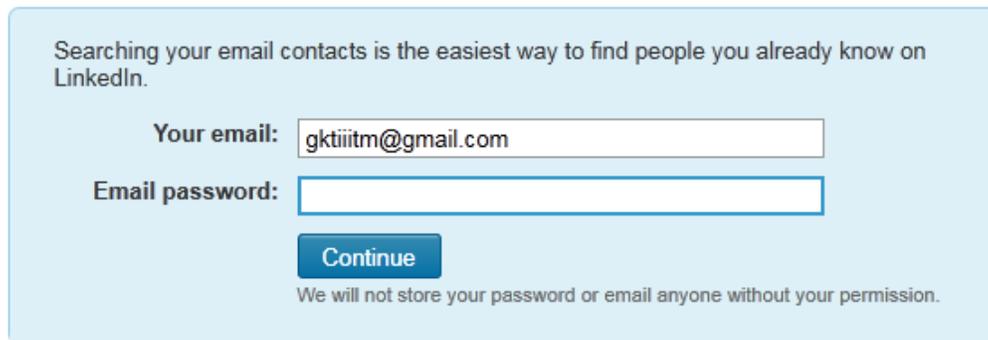

Figure 4. Functionality to invite friends on a social networking website

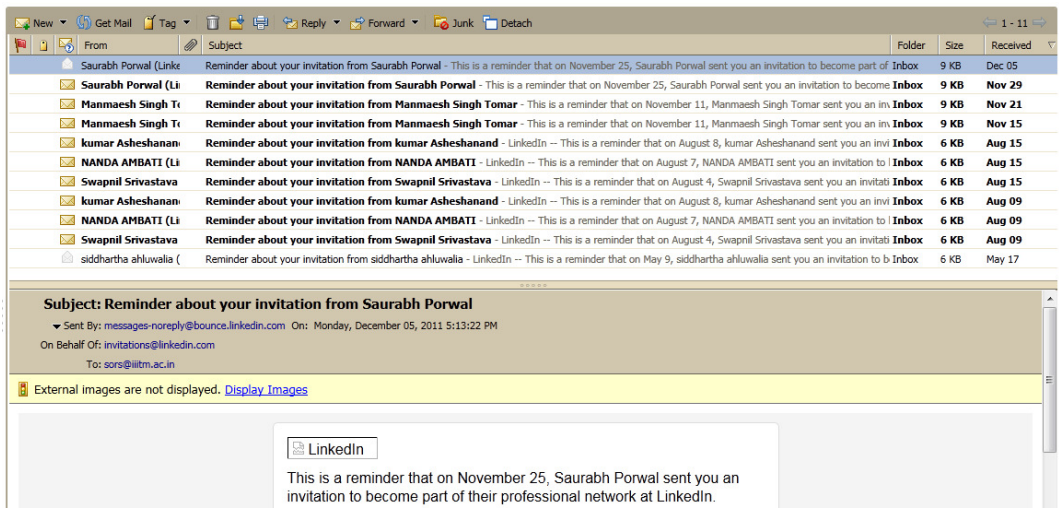

Figure 5. Invitation E-mails (spam) received as a result of activity in Fig. 4

Fig. 4 shows the feature on LinkedIn, a popular social networking website, to invite friends with an E-mail contact list. Fig. 5 shows the various invitation E-mail received on an official E-mail ID, sors@iiitm.ac.in, which is the E-mail address of the administrator of the Students' Online Record System (SORS), an online portal at Indian Institute of Information Technology and Management, Gwalior (IIITM), which manages information related to registered students of the university. It has been observed that SORS frequently receives many such invitation E-mails, advertisement E-mails, promotional E-mails and other kinds of spam.

This step in the proposed methodology suggests users to avoid giving their official E-mail address to untrusted people, advertisement groups or subscription lists. If this step is followed, E-mail users can protect themselves from unwanted E-mail attacks very easily, especially spam. They should simply give their personal/official E-mail IDs to only those whom they trust viz. family, close friends, etc.

### 3.2. Random URLs must not be clicked

Random URLs, especially those received in E-mail, posted on social networking or micro-blogging websites, or shortened ones which look unfamiliar, must not be clicked, Such URLs may not be trusted URLs and might be linking directly to pornographic sites or phishing sites, or they may be direct download links for viruses/malware.

As per our previous discussion, most of the malicious URLs are usually circulated through social networking websites, or E-mails, or even instant messaging sometimes. Links to obscene websites or phishing websites are spread in this way. Spam sent in the form of mass-mails usually gets detected by spam or phishing filters on E-mail servers but not all spams may be detected.

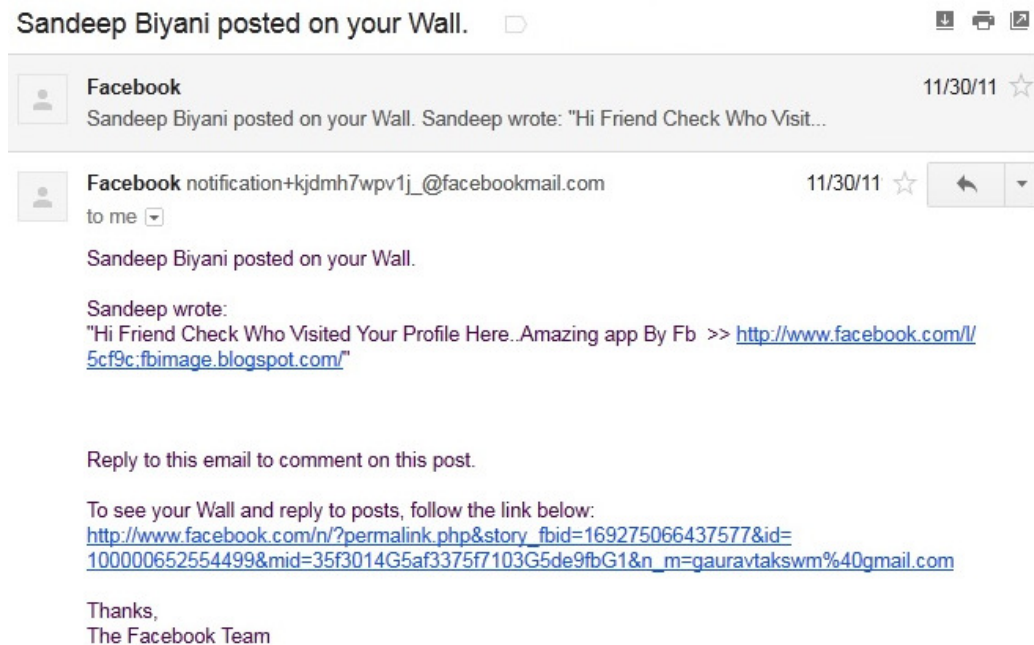

Figure 6. An E-mail notification from Facebook containing a malicious link

In Figure 6, a malicious URL is shown which has circulated through Facebook, and an E-mail notification has arrived at the mailbox. When an E-mail user clicks on this URL viz. http://www.facebook.com/l/5cf9c;fbimage.blogspot.com/ to know about the new activity, then "Hi Friend Check Who Visited Your Profile Here...Amazing app By Fb >> ", a porn link is spread out to all friends in the friend list without requesting any permission from the email user.

Sometimes, An online shopping firm sends a mail saying that there has been some problem in the billing information of the user and there is a need to login and verify the same. The included link re-directs to a site which is actually a phishing site and the login details are recorded in a database. There are many such attacks, which can be spread out with URLs by simply including them in E-mails

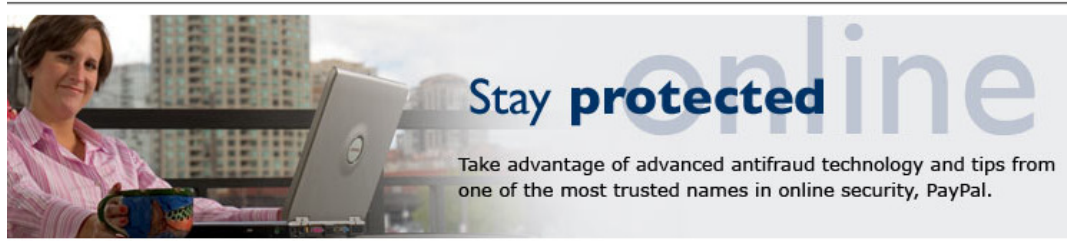

Figure 7. A PayPal phishing E-mail

In Figure 7, phishing E-mail of PayPal is shown. Any unsuspected user, who clicks on that link, might end up losing essential confidential information and may also have to face monetary losses. As per this step of the proposed methodology, E-mail users must be careful and avoid clicking on non-trusted URLs, as they may be phishing URLs. In this way, users may avoid the effects of most of the phishing-related and malicious mails.

### 3.3. Mails from untrustworthy people must not be responded to

Fig. 2 shows the structure of a Nigerian scam mail. Many a times, users receive E-mails related to winning cash prizes, lottery, gifts, or other kind of awards, which are completely bogus. Generally, these mails are related to unknown senders who just want to receive users' personal information, account information, or financial details for various fraudulent purposes.

As per this step in the proposed methodology, users are suggested not to respond to E-mails from unknown senders.

Whether an E-mail sender is known or unknown can be easily found out by the user, by simply looking at his/her previous conversations. For e.g., Gmail tries to fetch user information for every user on the network, so if user information is unavailable for an E-mail ID, it may be assumed that it is an unknown ID.

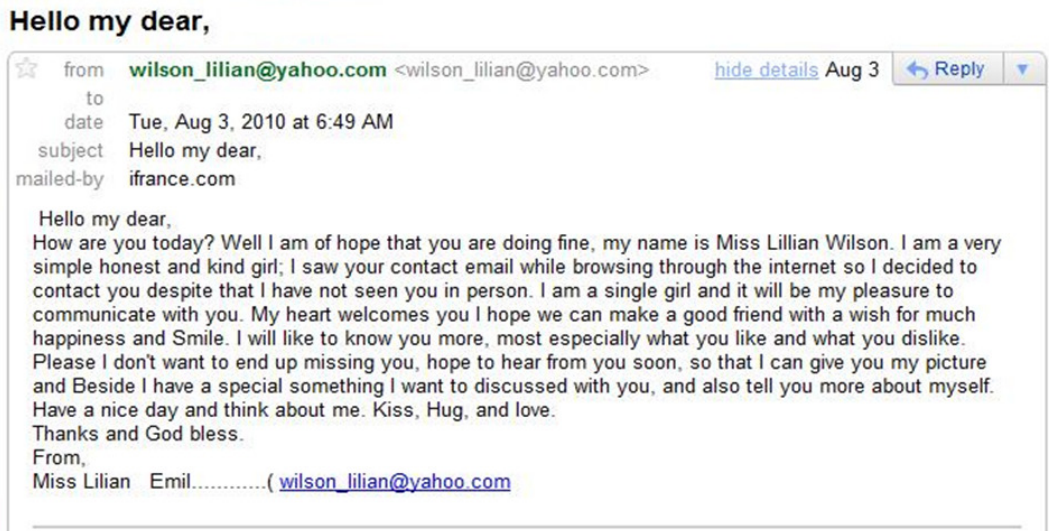

Figure 8. An E-mail from an unknown sender

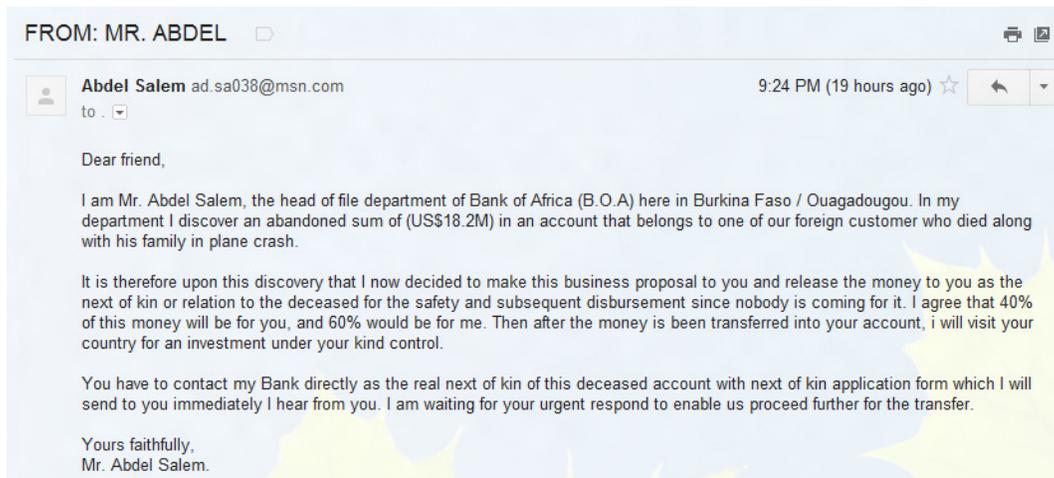

Figure 9. An E-mail from an untrustworthy sender (also scam)

As per this step in the proposed technique, users are suggested not to open any attachments from such mails, as they may be virus or malicious files, which may infect their computers' data or privacy.

### 3.4. Avoid using an official E-mail ID for registering or subscribing to mailing lists

At present, if users register on a social networking website using an E-mail ID, they get many notifications about their activity on such websites. This can be in the form of account activity, notifications, messages, or invitations. Due to the large proportion of such mails, users may miss their important or official E-mails. Sometimes, users receive excessive mails from subscriptions that have been willingly accepted.

Many social networking sites are known to sell their users' personal information such as contact details, social preferences and other information to advertisement agencies, which eventually fall in the list of spammers. Sometimes, E-mail notifications form an essential medium to spread viral infections on social networking sites.

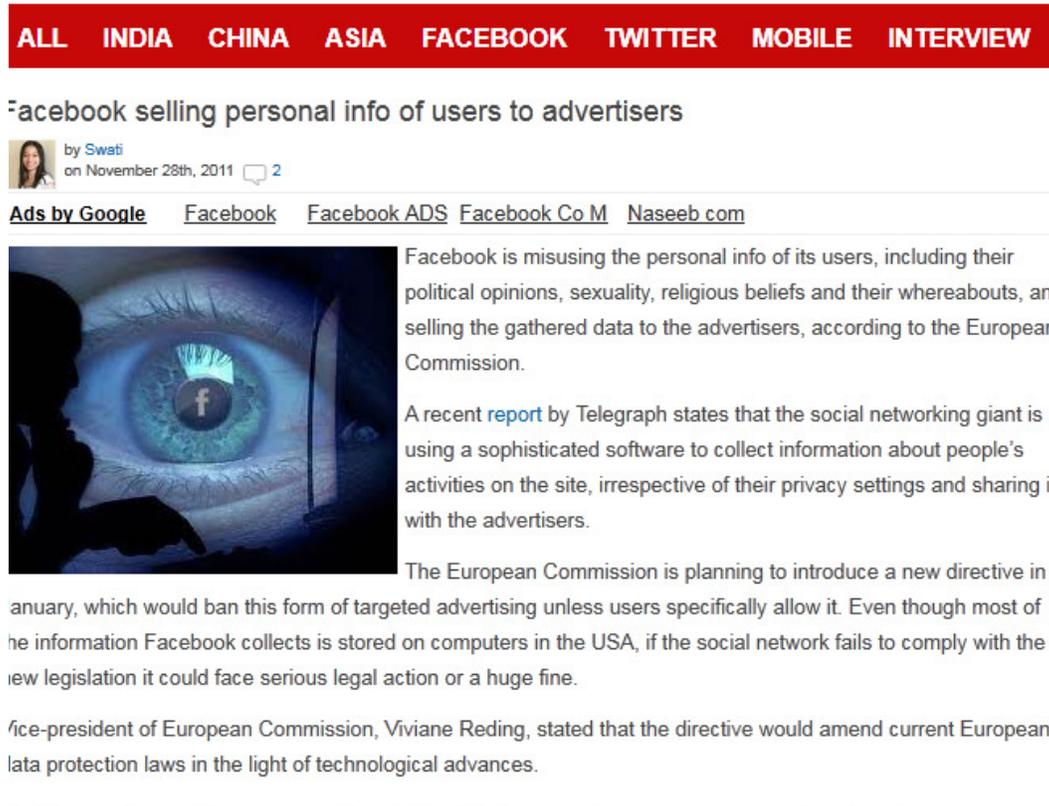

Figure 10. An article hinting that Facebook sells personal information to advertisers

Teenagers are the very keen when it comes to joining social networking websites, or dating sites. So much so that it has almost become a fashion. Some sites require premium membership to provide enhanced user experience, which requires the user to pay money. In this way, websites also acquire the financial details of their users. How they will use such details is outlined in the 'Trust and Privacy Policy' and 'Terms and conditions' which the users generally neglect. [link: http://www.buzzom.com/2011/11/facebook-selling-personal-info-of-users-to-advertisers/ ]. Some Facebook games give free credit if the user agrees to provide his/her credit card information to Facebook.

Using this step in the proposed methodology, E-mail users are suggested not to register on social networking sites with their official E-mail IDs, so that they can reduce the rate of E-mails (E-mails per day), thus managing their official mailboxes efficiently.

| | | |
|---|---|---|
| ☐ Facebook (2) | Stazy Khan commented on his photo of you. - facebook Hi Gaurav, Stazy Khan commented on | Dec 8 |
| ☐ Facebook (4) | Ambuj Tak commented on his photo of you. - facebook Hi Gaurav, Ambuj Tak commented on h | Dec 8 |
| ☐ Facebook | Ria Khandelwal commented on Ambuj Tak's photo of you. - facebook Hi Gaurav, Ria Khandel | Dec 8 |
| ☐ Facebook | Lotus Tak commented on Ambuj Tak's photo of you. - facebook Hi Gaurav, Lotus Tak commen | Dec 8 |
| ☐ Facebook | Amit Singh also commented on Kapil Kumar Mittal's status. - facebook Hi Gaurav, Amit Singh | Dec 8 |
| ☐ Facebook | Ajay Verma commented on Ankush Sharma's photo of you. - facebook Hi Gaurav, Ajay Verma | Dec 8 |
| ☐ Facebook | Mithilakishaan Abhishekanand also commented on Kapil Kumar Mittal's status. - facebook H | Dec 8 |
| ☐ Facebook | Deepak Verma also commented on Rakesh Kumar's status. - facebook Hi Gaurav, Deepak Ver | Dec 8 |
| ☐ Facebook | Vinu Thomas also commented on Sougat Kamilya's status. - facebook Hi Gaurav, Vinu Thoma | Dec 8 |
| ☐ Facebook | Abhishek DeathMetal also commented on Sougat Kamilya's status. - facebook Hi Gaurav, Ab | Dec 8 |
| ☐ Facebook | Aprajita Bhargava also commented on Heena Bhargava's photo. - facebook Hi Gaurav, Apraji | Dec 8 |
| ☐ Facebook | Ankush Sharma also commented on Shivika Sharma's status. - facebook Hi Gaurav, Ankush Shar | Dec 7 |

Figure 11. Notifications from a social networking website (Facebook)

## 3.5. Use special symbols while mentioning E-mail addresses on official websites

Now sometimes it happens that one needs to give one's E-mail ID on his/her official or company website for contact purposes. Some spammers use crawlers which are bots programmed to search for E-mail addresses on the web and create an index. Fig. 12 shows the approach used by a bot to extract E-mail addresses from text-based HTML pages

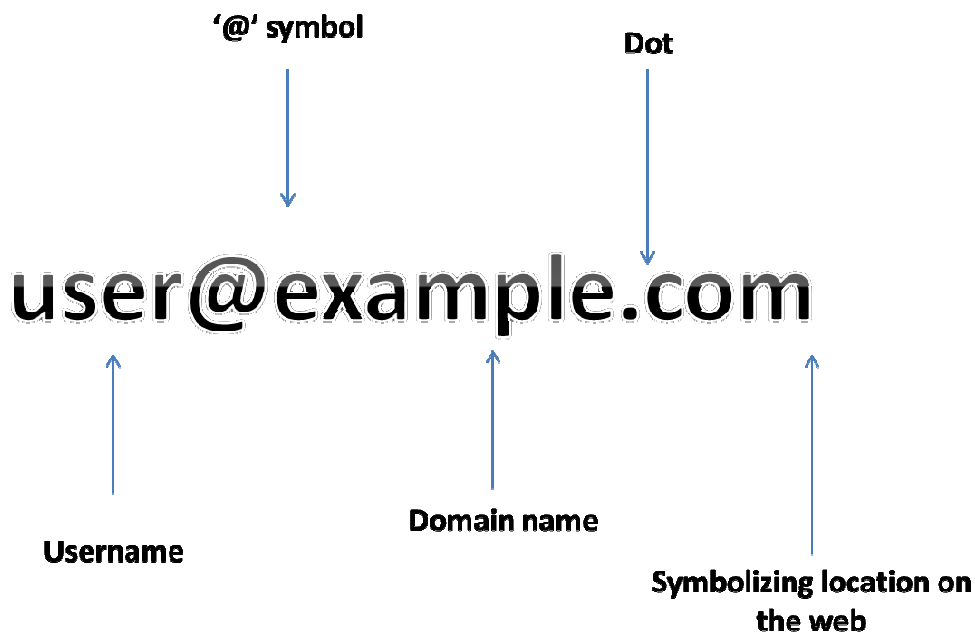

Figure 12. Splitting done by spammers' bots for searching E-mail addresses

Once the bot finds a string similar to the patterns defined in Fig. 12, it assumes that it is an E-mail address and stores it in its index.

As per this step in the proposed methodology, in order to counter this approach of the bot, one can simply be a little more careful while mentioning E-mail addresses on the web. Since the bot looks up for characters (@) and dots, one may write one's E-mail address in any of the following ways instead of writing it directly –

- object[@]domain.com
- object@domain[dot]com
- object[@]domain[dot]com
- object[at]domain[dot].com
- object AT domain DOT com
- OR similar other formats

In this way, bots will be unable to detect the presence of E-mail addresses but humans will always be able to interpret them. Thus, one can be sure that no spam or other related content will arrive in one's inbox.

## 4. ANALYSIS, EVALUATION AND RESULTS

If the steps in the proposed methodology are followed properly, users can avoid a large number of E-mail as well as other attacks. The approach is aimed at eliminating the step which makes these attacks meaningful viz. clicking a link, URL, or visiting a malicious site. Previously proposed methods are implemented at the server-side, and use complex algorithms, but the steps proposed in this paper are executed at the user's end. With the help of these steps, it is possible to identify nearly every kind of spam, scam, phishing mails, or malicious URLs. Depending upon the alertness of the user, this technique can give very low to zero false positives. The effectiveness of this approach goes on increasing with time as users learn to tackle various different kinds of attacks and get a general idea or pattern in these attacks. Not surprisingly, a famous proverb puts that '*Prevention is better than cure*'.

In this approach, although we have assumed this to be an independent one, the presence of background spam-detection methods in most mail services is not counted. If we take the general scenario then this method provides an extra layer of security thus increasing the efficiency very close to 100%.

To implement the approach and check results, we observed the mailboxes of twelve independent people, who knew very little about technical terms such as spams, scams or phishing, over a period of four months (March 2012 to June 2012). All the E-mail IDs were at least two years old with a few of them being the only IDs of their users. Most of them used Facebook, LinkedIn, Twitter, and associated services using those E-mail IDs which we analysed. Using the proposed steps, within a few days, the users were able to identify a majority of attacks. We also advised the users to not use the 'Mark as Spam' feature in Gmail, Yahoo!, and Hotmail services but separately move such mails to different folders as identified by them. The result is shown in Table 1, given below.

As per our results, the approach was 97.98% efficient overall. Personal as well as official mails were identified without any difficulty by the users, while advertisements were the most difficult to identify. This may be due to the complex nature of advertisements these days, whereby it is difficult to distinguish between useful advertisements (common advertisements) and useless advertisements (spam).

Table 1. Analysis of twelve different mailboxes over four months.

|  | Received Mails (A) | Identified Mails (B) | Efficiency in % [(B/A)* 100] |
|---|---|---|---|
| Total E-mails | 27001 | 26457 | 97.98 |
| Notifications | 5614 | 5598 | 99.71 |
| Spam/Scam | 7551 | 7467 | 98.89 |
| Phishing | 1662 | 1598 | 96.15 |
| Advertisements/Newsletters | 4840 | 4468 | 92.31 |
| Official mails/Receipts | 4901 | 4895 | 99.88 |
| Personal | 2433 | 2431 | 99.92 |

## 5. CONCLUSIONS

The proposed steps are highly effective in protecting E-mail users from unwanted E-mails of all kinds. Casual E-mail users could easily classify the various kinds of incoming mails and identify useless mails from the useful ones.

The results in table 1 show that the method is nearly 98% effective on an average. We reckon that the efficiency would only increase as users become more and more aware about the pattern in various kinds of E-mails. Also, if the built-in spam filters are used in addition to this technique, then we can achieve nearly 100% efficiency.

In future, we plan to implement all these steps are the server itself, with machine learning and artificial intelligence techniques. This will eliminate the need for the casual E-mail user to learn about various risks associated with spams, scams, phishing and other kinds of mails.

## ACKNOWLEDGEMENTS

We would like to thank everyone, especially students, friends, and family members who provided support and followed up with us for such a long period of four months, and trusted us regarding the content of their E-mails.

**Authors**

**Gaurav Kumar Tak** is an Assistant Professor in CSE Dept., School of Computer Science and Information Technology, Lovely Professional University, Phagwara, Punjab. He earned his M.Tech and B.Tech in ICT (Information Communication Technology) from ABV-Indian Institute of Information Technology and Management, Gwalior, India in 2011. His primary research areas of interest are Computer Networks, Adhoc Networks, Information Security, Cyber Crimes and Security, etc.

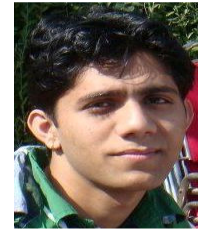

**Gaurav Ojha** is a student in the Department of Information Technology at Indian Institute of Information Technology and Management, Gwalior. His areas of interest are Web Technologies, Open Source Software and Internet Security.

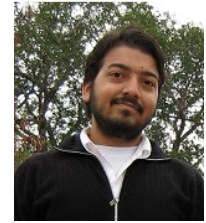